\documentclass[twocolumn,amsmath,amssymb,superscriptaddress]{revtex4}

\usepackage{graphicx}
\usepackage{dcolumn}
\usepackage{bm}

\begin{document}


\title{A multi-purpose modular system for high-resolution microscopy at high hydrostatic pressure
}

\author{Hugh Vass$^*$}
 \affiliation{SUPA, School of Physics and Astronomy, University of Edinburgh, James Clerk Maxwell Building, The King's Buildings, Mayfield Road, Edinburgh EH9 3JZ, UK}

\thanks{equal contribution, ** equal contribution}

\author{S. Lucas Black$^*$}%
 \affiliation{SUPA, School of Physics and Astronomy, University of Edinburgh, James Clerk Maxwell Building, The King's Buildings, Mayfield Road, Edinburgh EH9 3JZ, UK}
\affiliation{Institute of Cell Biology, School of Biological Sciences, University of Edinburgh, Darwin Building, Mayfield Road, Edinburgh EH9 3JR, UK}
\author{Eva M. Herzig$^*$}
\affiliation{SUPA, School of Physics and Astronomy, University of Edinburgh, James Clerk Maxwell Building, The King's Buildings, Mayfield Road, Edinburgh EH9 3JZ, UK}
\author{F. Bruce Ward$^{**}$}
\affiliation{Institute of Cell Biology, School of Biological Sciences, University of Edinburgh, Darwin Building, Mayfield Road, Edinburgh EH9 3JR, UK}
\author{Paul S. Clegg$^{**}$}
\affiliation{SUPA, School of Physics and Astronomy, University of Edinburgh, James Clerk Maxwell Building, The King's Buildings, Mayfield Road, Edinburgh EH9 3JZ, UK}
\author{Rosalind J. Allen$^{**}$}
\affiliation{SUPA, School of Physics and Astronomy, University of Edinburgh, James Clerk Maxwell Building, The King's Buildings, Mayfield Road, Edinburgh EH9 3JZ, UK}
\email{rallen2@ph.ed.ac.uk}

\date{\today}

\begin{abstract}
We have developed a modular system for high-resolution microscopy at high hydrostatic pressure.   The system consists of a pressurised cell of volume $\sim$100 $\mu$l, a temperature controlled holder, a ram and a piston. We have made each of these components in several versions which can be interchanged to allow a wide range of applications. Here, we report two pressure cells with pressure ranges 0.1-700MPa and 0.1-100MPa, which can be combined with hollow or solid rams and pistons. Our system is designed to work with fluorescent samples (using a confocal or epifluorescence microscope), but also allows for transmitted light microscopy via the hollow ram and piston. The system allows precise control of pressure  and temperature [-20-70$^\circ$C], as well as rapid pressure quenching. We demonstrate its performance and versatility with two applications: time-resolved imaging of colloidal phase transitions caused by pressure changes between 0.1MPa and 101MPa, and imaging the growth of {\em{Escherichia coli}} bacteria at 50MPa.  We also show that the isotropic-nematic phase transition of pentyl-cyanobiphenyl (5CB) liquid crystal provides a simple, convenient and accurate method for calibrating pressure in the range 0.1-200MPa.  
\end{abstract}

\maketitle

\section{Introduction}\label{sec:intro}

Hydrostatic pressure is important in many research fields. The structural, physical and chemical properties of matter under pressure are of interest to physicists, chemists and materials scientists \cite{mcmahon,pae,deguchi2}. Hydrostatic pressure is also an important tool in elucidating the mechanisms underlying the assembly of protein and other macromolecular structures \cite{smeller,macgregor,winter,heremans}, while pressure effects on cartilage and bone cells are physiologically and medically important \cite{urban}. From a technological point of view, moderate hydrostatic pressures ($\sim$ 200MPa) have potential applications in protein manufacturing \cite{seefeldt}, while treatment with higher hydrostatic pressures  ($>$ 600MPa) can be used to kill microorganisms in food sterilization \cite{martin} or surgical applications \cite{diehl}. Finally, deep sea environments, which constitute a large part of the global biosphere, are exposed to hydrostatic pressures up to 101MPa. Understanding how biological organisms adapt to such pressures is of both fundamental scientific and biotechnological interest  \cite{bartlett,gross,abe,kato}. 

In all of these contexts, optical microscopy is a powerful tool. Since depressurization before imaging constitutes a large and uncontrolled perturbation, it is a great advantage to be able to image a pressurized sample {\em{in situ}} at elevated pressure. However, microscope pressure cells are expensive and time-consuming to construct, and are often designed for a single application, making it difficult to use a particular system for multiple applications. In this paper, we report the development of a multipurpose microscope pressure-cell system, in which different modules can be interchanged to allow for different requirements in terms of microscopy technique, solvent tolerance, optical resolution and pressure range. The system can be used with a conventional epifluorescence, transmission or confocal microscope, has a large sample volume of $\sim$100$\mu$l, is temperature controlled and is capable of rapid and well-controlled pressure ramping or quenching. Our focus is on biological and soft-matter physics applications in the pressure range $0.1-700$MPa. 

A microscope pressure cell suitable for biological and soft-matter applications must have a large sample volume ($\sim$ 100$\mu$l), in order to be able to observe ``bulk-like'' material or cellular behaviour. The pressure cell must also be compact and light enough to fit on a standard microscope stage. The cell must have optical windows which can sustain pressure. However, good optical resolution requires the working distance between the objective lens and the sample to be as small as possible. The window should therefore be as thin as possible and the window aperture as wide as possible to allow close approach of the objective to the window. One would also like to achieve well-controlled and rapid pressure quenching and temperature control.  A number of pressure cells have been reported for imaging at moderate hydrostatic pressure; the properties of a representative sample of these are summarized in Table \ref{tab:cells}. At higher pressures (several GPa), diamond anvil cells \cite{jayamaran} are widely used; their small sample volume and thick diamond windows make them generally unsuitable for high resolution imaging of biological and soft matter samples under moderate pressures, although modified diamond anvil cells for imaging biological samples of small volume have been developed \cite{oger}.  

\begin{table*}
{\scriptsize \begin{tabular}{l|l|l|l|l|l|l}
Ref.&Application&Sample &Max&Optical properties &Pressurization& Special features\\
&&volume &Pressure&&& \\
\hline
\cite{oger} & Microbiology & 0.1$\mu$l & 1.4GPa & Diamond window& Gas-filled ram & Diamond anvil cell\\
 &  & &  &250-600$\mu$m thick, 4mm diam.   &  & Temp range to 300$^\circ$C \\
 &  & &  &  20$\times$ objective &  &  \\
\hline
\cite{perriercornet} & Microbiology & 25$\mu$l & 700MPa & Sapphire window& Hand pump & \\
 &  & &  &5mm thick, 10mm diam.   &  &  \\
 &  & &  &  20$\times$ objective &  &  \\
\hline
\cite{reck} & Polymer physics  & $\sim$300$\mu$l & 300MPa & sapphire or diamond window & Hand pump & Polarization microscopy  \\
 &  & &  &$\sim$2mm thick, $\sim$ 5mm diam, WD 12mm &  &  Temp range -40-270$^\circ$C \\
 &  & &  & 10$\times$ objective &  \\
\hline
\cite{maeda} & Liquid crystals & &  300MPa & Sapphire window&  Hand pump & Temp range 20-250$^\circ$C \\
 & Polymer physics & &  &5mm diam., WD 3.5mm &  \\
 &  & &  & 20$\times$ objective & \\
 \hline
\cite{hartmann,hartmann2,frey} & Cell Biology  & 3.5$\mu$l & 300MPa & Sapphire window & Gasket  & Fluorescence and transmitted\\
 &  & &  &2.3mm thick, WD 1.5mm. &  \\
&  & &  &  20$\times$ or 40$\times$ objective &  \\
\hline
\cite{nishiyama} & Biophysics & 200 $\mu$l& 200MPa & Optical glass window & Hand pump& \\
 &  & &  &1.5mm diam.,  WD 6mm  &  & \\
 &  & &  &  40$\times$ objective &  &   \\
\hline
\cite{koyama}& Cell biology & 100$\mu$l &  100MPa &  Pyrex glass window,  & HPLC pump & Continuous flow \\
&  & &  &2mm thick, 2mm diam., WD 3.8mm  &  & Temp range 2-80$^\circ$C\\
 &  & &  &  20$\times$ or 40$\times$  objective & \\
\hline
\cite{salmon} & Cell Biology & 75$\mu$l &   80MPa & Strain-free glass&Hand pump & Polarization microscopy\\
 &  & &  &1.75mm thick, 3mm diam., WD 2.3mm & \\
 &  & &  & 40X objective & \\
\hline
\cite{raber} & Biological cells  & $\sim $1$\mu$l  & 100MPa  & Image through wall of glass capillary & Hand pump &Fluorescence microscopy\\
 &  & &  &0.34 or 0.5mm thickness &  &Glass capillary tube \\
 &  & &  & 10X objective &  \\
\hline
\cite{deguchi,mukai} & Colloid physics & 190$\mu$l& 40MPa & Diamond window &  HPLC pump & Continuous flow \\
 &  & &  &1mm thick, 2.8mm diam, NA 0.25 &  &  Temp range 20-450$^\circ$C\\
 &  & &  & 10$\times$ objective & \\
\hline
\cite{besch} & Cell Biology & 100$\mu$l & 15MPa &  Glass coverslip & HPLC & Continuous flow \\
 &  & &  &0.15mm thick,  1mm diam. &  & Fluorescence and transmitted  \\
 &  & &  &  40$\times$ objective &  & Electrical stimulation of cells \\
\hline
\cite{mizuno} &  Cell Biology & 9ml & 10MPa  & Sapphire window & HPLC flow cell  & \\
 &  & &  &2mm thick, 30mm diam. & \\
 &  & &  & 40$\times$ objective &  \\
\hline
\cite{pagliaro} & Cell Biology & $\sim $1ml & 7MPa &Glass coverslip&  HPLC pump & Fluorescence and transmitted\\
 &  & &  &0.2mm thick, 1mm diam.  & \\
 &  & &  & 40$\times$, 1.3NA objective (fluorescence) &  \\
\end{tabular}}
\caption{Technical properties of a representative sample of high-pressure microscope cells  for biological and soft-matter applications, as reported in the literature.\label{tab:cells}}
\end{table*}

Although the cells listed in Table \ref{tab:cells} represent a wide range of capabilities, each is designed for a specific application. Our aim was to develop a modular and multipurpose system, allowing for high-resolution high-pressure microscopy over a range of pressures up to ~700MPa, suitable for users in a variety of research fields. Given the expense and technical difficulty associated with the construction of a pressure cell, such a multi-purpose system should significantly facilitate high-pressure research. In Section \ref{sec:des} of this paper, we describe our system in detail. In Section \ref{sec:cal} we discuss calibration and optical testing. In Sections \ref{sec:colloids} and \ref{sec:bacteria}, we demonstrate the system by presenting data on the kinetics of a pressure quench-induced colloidal phase transition and high-resolution imaging of bacteria growing under pressure. Finally, we present our conclusions in Section \ref{sec:concs}.

\section{Pressure cell design}\label{sec:des}

\begin{figure}[h!]
\begin{center}
\makebox[20pt][l]{(a)}{\rotatebox{0}{{\includegraphics[scale=0.3,clip=true]{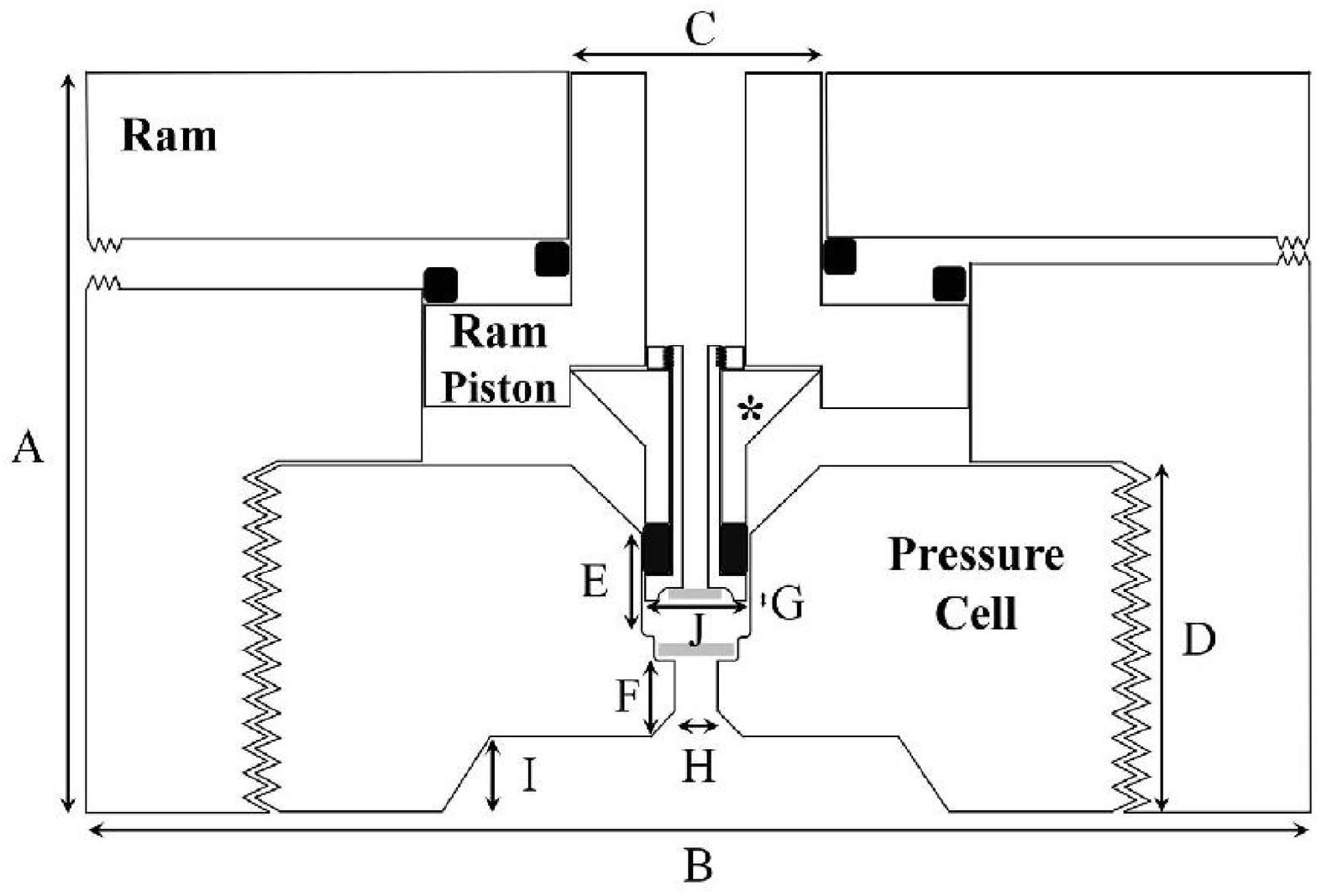}}}}\\\vspace{0.25cm}\makebox[20pt][l]{(b)}{\rotatebox{0}{{\includegraphics[scale=1.0,clip=true]{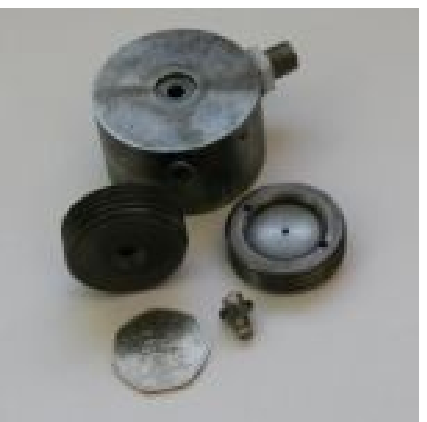}}}}
\caption{(a): Cross-section of the pressure cell mounted in the pressurizing ram. The star denotes the  pressure cell piston. The black squares indicate seals. The dimensions labelled by letters are listed in Table \ref{tab:dims}. (b): A photograph showing the hollow ram, pressure cells 1 and 2 and the hollow pressure cell piston, with coin (diameter 2.7cm) for size comparison. \label{fig:diag}}
\end{center}
\end{figure}

Figures \ref{fig:diag}(a) illustrates the design of our pressure cell system, a photograph of which is shown in Figure \ref{fig:diag}(b). Dimensions are given in Table \ref{tab:dims}. The system consists of two basic parts: the pressure cell and the pressurizing ram. Each of these parts has an associated piston. The pressure cell, of diameter 35mm, contains the sample chamber [diameter 4.5mm], which has a window aperture of diameter 1.5mm. A piston of diameter 4.5mm fits into the cell. The pressure cell is screwed into the ram, in such a way that the ram piston exerts pressure on the pressure-cell piston when hydraulic fluid is pumped into the ram. Brake fluid is used as the hydraulic fluid of choice, particularly for rapid quenching, due to its low compressibility and temperature-independent properties.

\begin{table}
\begin{tabular}{l|c|c|c}
\multicolumn{4}{c}{Rams}\\
&A&B&C\\
\hline
Solid ram&\,29.5\,&\,50.0\,&\,0.0\,\\
Hollow ram\,\,&29.5&50.0&10.0\\
\end{tabular}\vspace{0.1cm}\\
\begin{tabular}{l|c|c|c|c|c|c}
\multicolumn{7}{c}{Pressure cells}\\
&D&E&F & G & H& I\\
\hline
Cell 1&\,13.5\,&\,3.0\,&\,4.0\,&\,0.5\, &\, 1.5\, &\, 3.0\\
Cell 2&10.5&3.5&1.5&0.5 & 1.5 & 2.5\\
\end{tabular}\vspace{0.1cm}\\
\begin{tabular}{l|c|c}
\multicolumn{2}{c}{Pressure cell pistons}\\
&J\\
\hline
Solid piston &\,4.5\\
Hollow piston &4.5\\
\end{tabular}
\caption{Dimensions of the rams, pressure cells and pistons. The letters refer to the illustration in Figure \ref{fig:diag}. All dimensions are in millimetres.\label{tab:dims}}
\end{table}

The pressure cell is made of TiMetal 550 titanium alloy, which is machined to shape and heat treated for maximum tensile strength. This heat treatment also causes the metal surface to darken, reducing the reflection of light inside the cell. The pressure cell window is made of either 0.5mm thick type 2A gem quality diamond (Element 6, Cuijk, The Netherlands) or 0.45mm optical quality quartz (UQG Optics). The window is attached to the optically flat surface by a Poulter-type seal and fixed at 180$^\circ$C with chemically resistant adhesive (MBond). The pressure cell piston is made of TiMetal 550, machined and heat treated as described above. Sealing rings are made of polytetrafluoroethylene (PTFE) for low pressure sealing (up to 200MPa) and phosphorbronze plus PTFE for high pressure sealing (up to 700MPa). For transmitted light microscopy, we have made a version of the pressure cell piston which is hollow and contains a 0.5mm thick diamond window, fixed using a Poulter-type seal as described above. Light can pass into the cell through this window, when used in conjunction with a hollow version of the ram piston. The ram is made of w-720 tool steel (B{\"{o}}hler UK), fitted to standard hydraulic connectors, and the ram piston uses single or dual O-ring seals (solid and hollow versions respectively). 

We have constructed multiple versions of both the ram and the pressure cell. These have different functionalities and can be combined in a modular fashion. The hollow version of the ram contains a hollow piston, and is suitable for transmitted light microscopy. An alternative, solid, ram contains instead a solid piston. The solid ram has a simpler and more rugged design, making it more suited to higher pressures. Pressure cell 1 is designed for pressures up to 700MPa and has a diamond window  of thickness 0.5mm. This cell allows for a working distance of 4mm and is compatible with a commercial 20$\times$ Extra Long Working Distance (ELWD) objective (Nikon; working distance: 8.1-7.0mm, NA: 0.45). Pressure cell 2 is designed for higher resolution at pressures up to ~100MPa. This cell has a quartz window of thickness 0.45mm, with a more open window surround, and can achieve a working distance of 1.4mm, making it compatible with a 60$\times$ ELWD objective (Nikon; working distance: 1.5-2.1mm, NA: 0.70). The pressure cells can be used in combination with either of two pressure cell pistons: a hollow piston containing a diamond window, which is suitable for transmitted light imaging, and a more rugged solid piston.

\section{Calibration and testing}\label{sec:cal}

\subsection{Pressure calibration}

\subsubsection{Higher pressure (200-700MPa)}

\begin{figure}[h!]
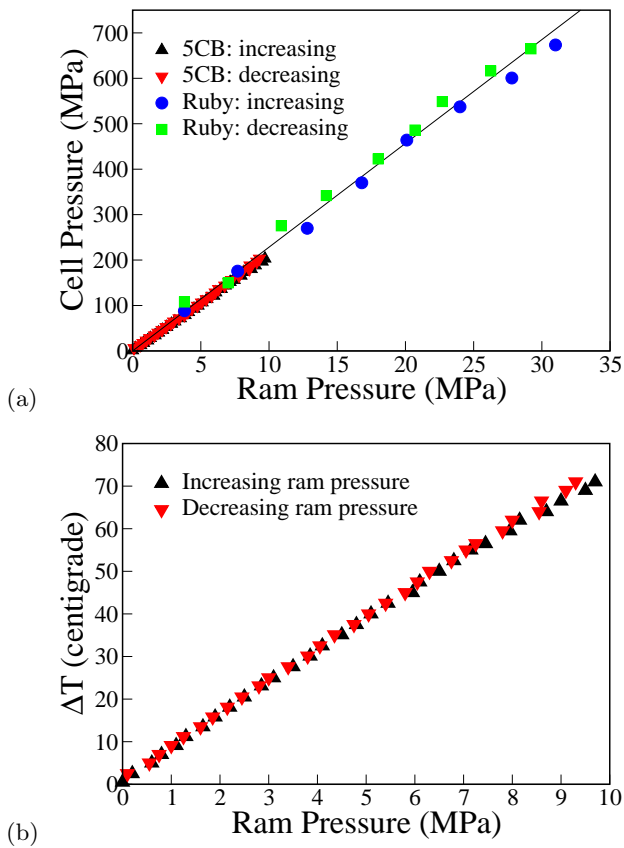

\begin{center}
\makebox[20pt][l]{(a)}{\rotatebox{0}{{\includegraphics[scale=0.3,clip=true]{fig2a.eps}}}}\\\vspace{0.25cm}\makebox[20pt][l]{(b)}{\rotatebox{0}{{\includegraphics[scale=0.3,clip=true]{fig2b.eps}}}}
\caption{Calibration data for pressure cell 1 with the hollow ram. (a): Cell pressure as a function of ram pressure for the ruby calibration and for the 5CB calibration (using the data from panel (b)). The solid line indicates the ``ideal'' results if the seals were 100\% efficient and frictionless (calculated from the ratio of piston areas, assuming the ram pressure gauge to be accurate). Blue circles represent ruby data collected for increasing pressure, while green squares denote ruby data for decreasing pressure. Upward black triangles: 5CB data for increasing ram pressure; downward red triangles: 5CB data for decreasing ram pressure.(b): 5CB transition temperature as a function of ram pressure ($\Delta T$ refers to the temperature above the ambient pressure transition). Symbols are as in panel (a). \label{fig:cal}}
\end{center}
\end{figure}

In the higher pressure range (200-700MPa), the ruby fluorescence technique (R1 line) was used for calibration \cite{grasset}. A tiny crystal of natural ruby was attached to the inside of the diamond window with transparent araldite and the cell was filled with water. Temperature was controlled at 26.0$\pm$0.1$^\circ$C. A neon bulb was used to provide a calibration reference line, in near coincidence with the R2 ruby line. A Coderg triple 800 spectrometer set for high resolution (0.33cm$^{-1}$), with steps of 0.125cm$^{-1}$, was used to obtain ruby spectra at different pressures, and the best fit to a Lorentzian lineshape  was used to obtain the position of the R1 ruby line, relative to the neon reference, allowing the pressure inside the cell to be determined. A high resolution spectrometer is required because the magnitude of the R1 shift is very small. Results for pressure cell 1 are shown in Figure \ref{fig:cal}(a). The calibration was done with an Enarpac Bourdon gauge with a scale reading to 40 MPa, with the fluorescence wavelength shift taken as 7.68 cm$^{-1}$ / GPa.

\subsubsection{Lower pressure ($<$200MPa)}

In the lower pressure range (below 200MPa), a liquid crystal calibration method was used. This provides a remarkably simple yet accurate method of calibration which requires only accurate temperature control. The method involves observing the nematic to isotropic phase transition of pentyl-cyanobiphenyl (5CB) as the cell is heated in stages through the transition into the isotropic phase. The transition pressure in the cell can then be calculated from the formula derived by Shashidhar and Ventkatesh \cite{shashidar} for 5CB: 
\begin{equation}\label{eq:sv}
\Delta T=40.3P-2.64P^2
\end{equation}
 where $\Delta T$ is the temperature above the ambient pressure transition temperature (in centigrade) and P is in kbar (1kbar$=$101MPa). Ram pressure is then gradually applied until the nematic phase is restored. By slowly reducing the ram pressure until the transition reverses, a measurement of the hysteresis of the cell at that pressure is obtained.

To observe the transition, about 2 to 3 mm depth of neat 5CB was placed in the cell and sealed with the light transmitting (hollow) pressure cell piston. A small amount of laser light passed through the cell and was observed on a screen. As pressure was applied, the laser spot shimmered just before the transition before being abruptly extinguished at the transition. After about 30 seconds a much weaker spot appeared on the screen with a polarisation generally different to the laser polarisation. On slowly releasing the ram pressure, the spot shimmered and then became much brighter as the transition reversed.

Although a laser was used in this calibration, it is not essential: the transition can be observed quite easily with ordinary light. As the cell pressure is derived directly from the temperature, a cheap low accuracy Bourdon gauge is sufficient for the ram pressure measurement, as long as the calibration curve is re-measured for each gauge.

Figure \ref{fig:cal}(b) shows a 5CB calibration of pressure cell 1.  The ram pressure was measured with a small 10MPa R.S. Bourdon gauge. Figure \ref{fig:cal}(a) (triangles) shows the same data, with the vertical axis translated into the cell pressure using Eq.(\ref{eq:sv}).  The solid line shows the expected pressure in the cell if the seals were 100\% efficient and had zero friction. Both the 5CB and ruby calibrations in Figure \ref{fig:cal}(a) show straight lines with very little hysteresis. The gradient of the line for the 5CB calibration is slightly lower than that for the ruby calibration, suggesting that the 10MPa gauge  was slightly overestimating the ram pressure with respect to the 40MPa gauge. This was later confirmed in a separate measurement. We note that the 5CB calibration method could be extended to higher pressures for pressure cells which can withstand very high temperatures.

\subsection{Optical resolution}
To evaluate the optical resolution of our two pressure cells, we imaged 3$\mu$m fluorescent beads (Sigma-Aldrich) in a Nikon/Bio-Rad confocal microscope with a 488nm laser. Figure 3 shows horizontal (a and b) and vertical (c and d) section images of a single bead, averaged over 8 frames and corrected for the sample medium \cite{white}. While we are able to resolve single beads in both pressure cells, significantly better resolution is achieved with cell 2. 

\begin{figure}[h!]
\begin{center}
\makebox[20pt][l]{(a)}{\rotatebox{0}{{\includegraphics[scale=0.4,clip=true]{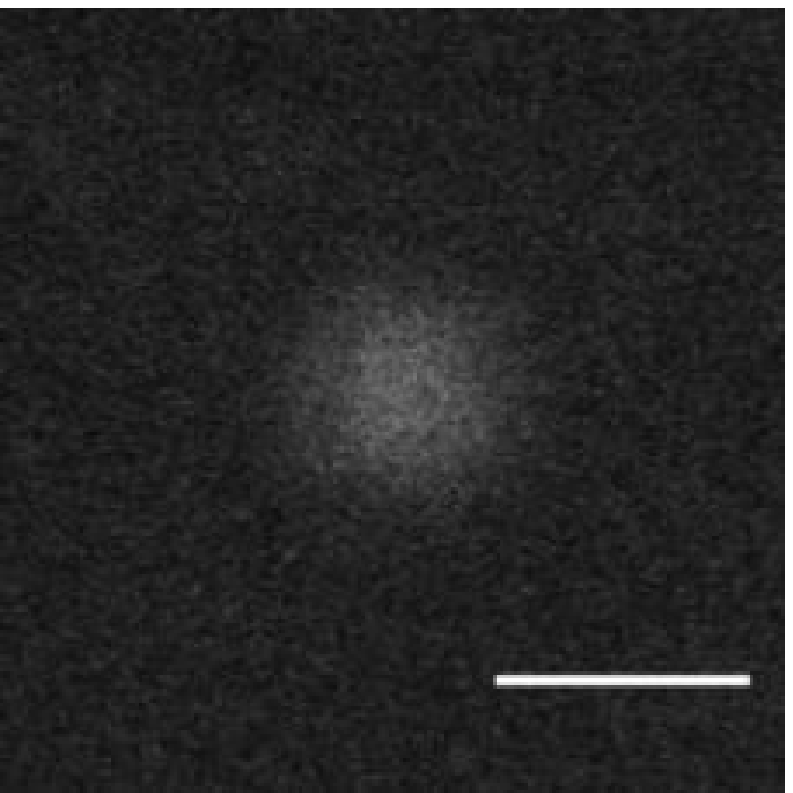}}}}\hspace{0.25cm}\makebox[20pt][l]{(b)}{\rotatebox{0}{{\includegraphics[scale=0.131,clip=true]{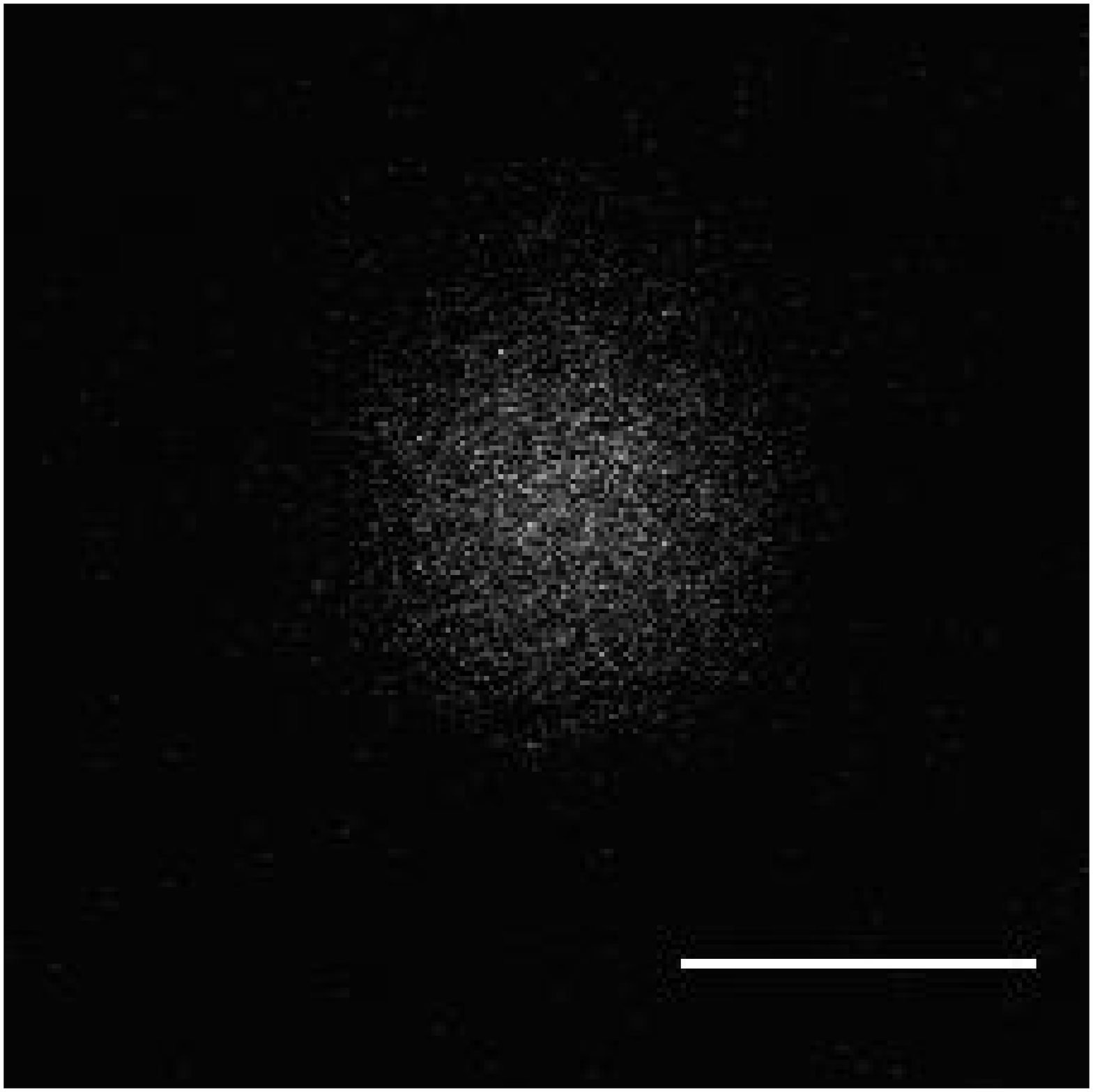}}}}\\\makebox[20pt][l]{(c)}{\rotatebox{0}{{\includegraphics[scale=0.4,clip=true]{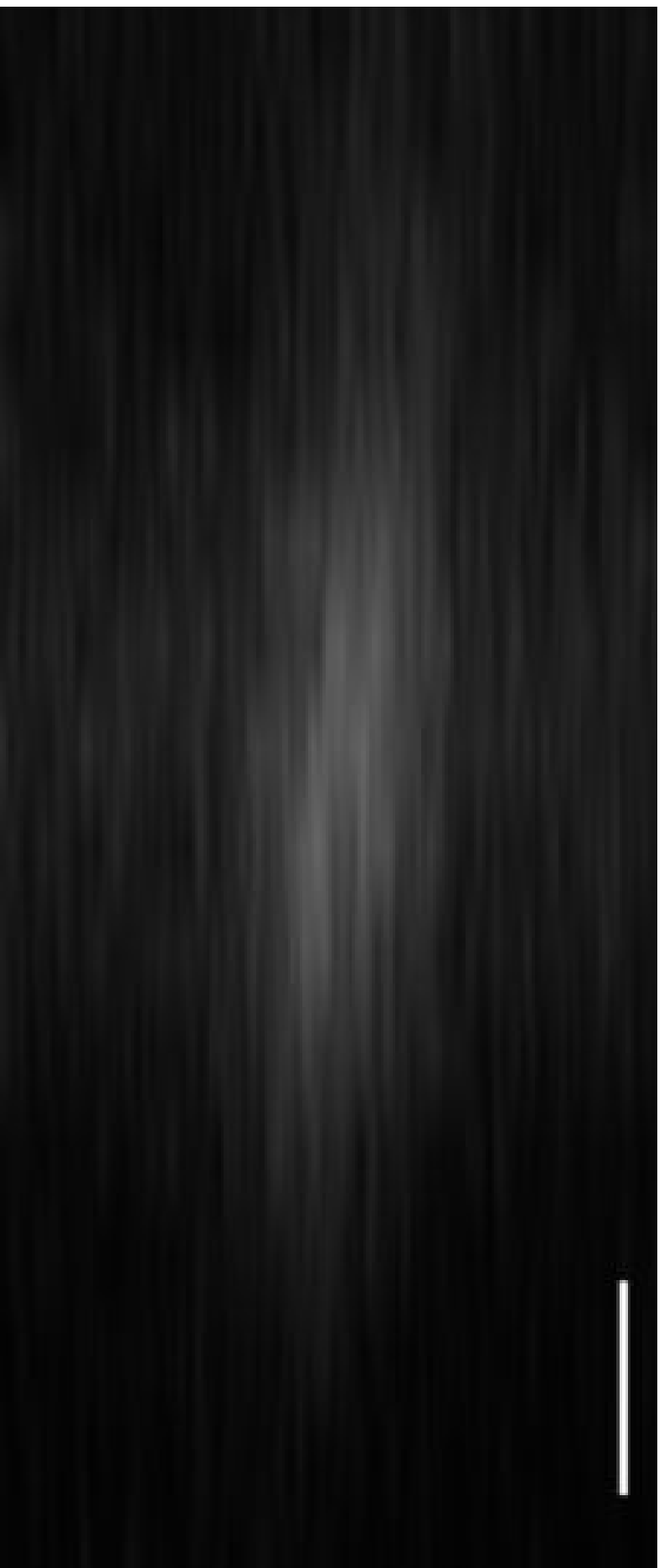}}}}\hspace{0.25cm}\makebox[20pt][l]{(d)}{\rotatebox{0}{{\includegraphics[scale=0.131,clip=true]{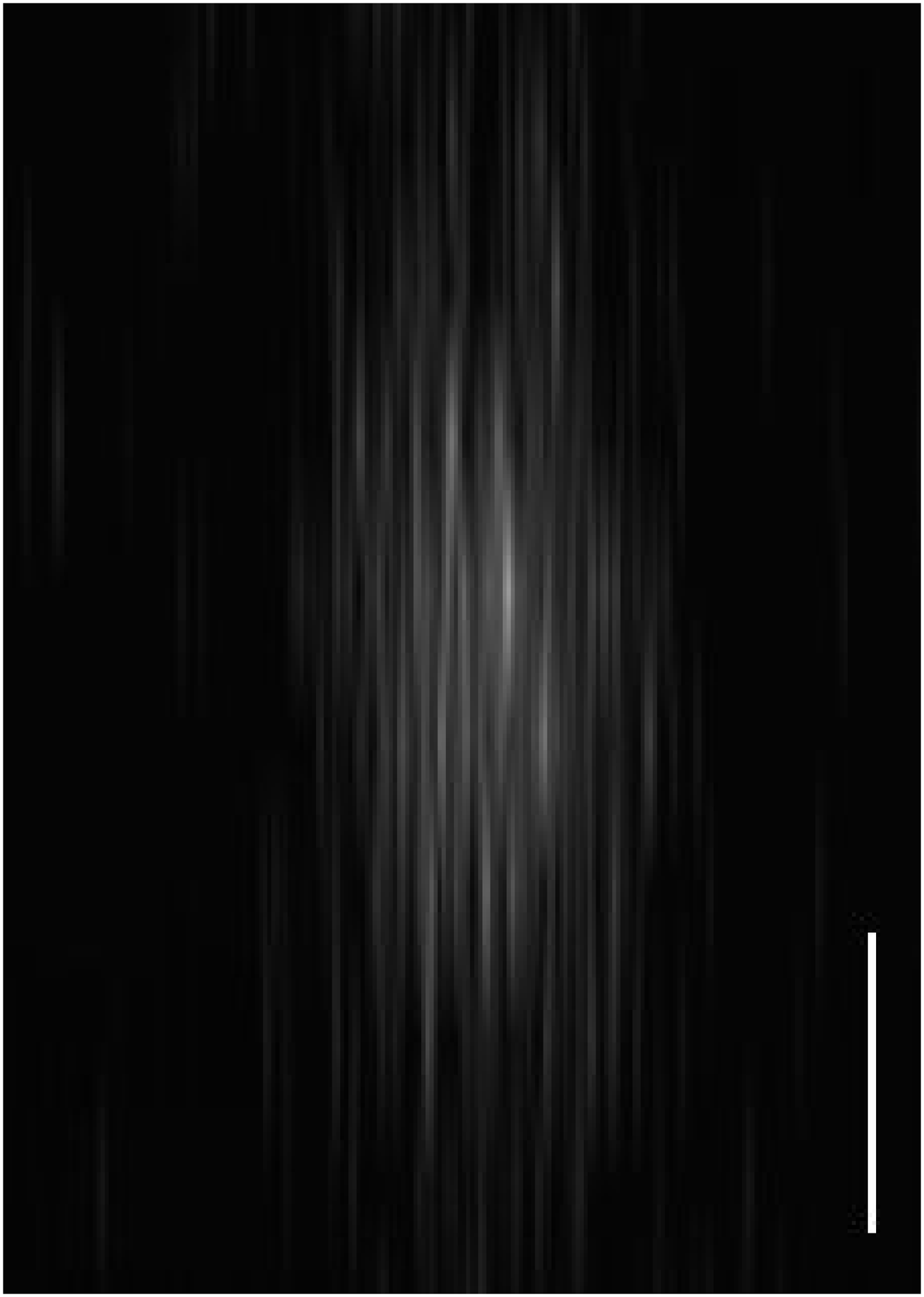}}}}
\caption{Images of a 3$\mu$m fluorescent bead, obtained in a Nikon / Bio-Rad confocal microscope and corrected for the sample medium \cite{white}. (a) x-y plane image taken in pressure cell 1. (b) x-y plane image taken in pressure cell 2.  (a) x-z plane image taken in pressure cell 1. (d) x-z plane image taken in pressure cell 2. The scale bar is 3$\mu$m in each image.\label{fig:opt}}
\end{center}
\end{figure}

\section{Application 1: Colloidal phase transition kinetics}\label{sec:colloids}
Hydrostatic pressure can be used in colloid science to modify interactions, change the relative composition of multi-component samples (if the components have different compressibilities) and to reach new nonequilibrium states by rapid quenching. For example, the effect of pressure on hydrophobic interactions has been explored experimentally for proteins \cite{meersman2006} and polymers \cite{meersman2005} and via simulations for proteins and hydrophobic solutes \cite{hummer,hillson,ghosh,ghosh2002}. Other researchers have used hydrostatic pressure to investigate the phase diagram of a system of adhesive colloids \cite{vavrin} and to adjust the volume fraction of a sample of colloidal rods, leading to a transition from the isotropic phase into the nematic phase \cite{lettinga}. These studies involved slow changes of pressure, but it is possible to achieve pressure changes at the speed of sound, significantly faster than changes of temperature. Pressure therefore presents promising possibilities for experiments requiring fast quench rates. For example, the kinetics of relaxation to equilibrium in binary fluids have been investigated using rapid pressure quenches combined with light scattering  \cite{wong}. Our interest is in the spatial structure  of binary fluids undergoing a phase separation which is arrested by the jamming of colloidal particles at the fluid interfaces \cite{herzig}. These systems are highly heterogeneous, making  fluorescence confocal microscopy the technique of choice. Here, we present data obtained using pressure cell 1 (which has a 0.5mm thick diamond window), in combination with fluorescence confocal microscopy with a 20$\times$ ELWD objective, for both slow and rapid pressure quenches.

\begin{figure}[h!]
\begin{center}
{\rotatebox{0}{{\includegraphics[scale=1.0,clip=true]{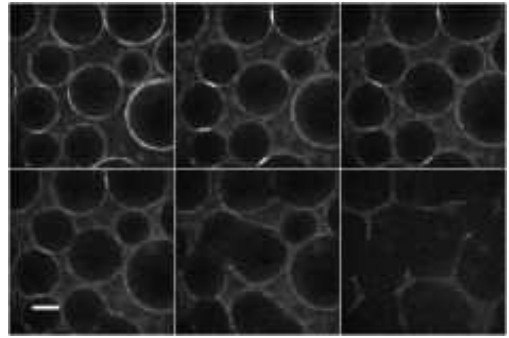}}}}
\caption{Fluorescence confocal microscopy images of a phase transition from immiscible to miscible states by applying a slow pressure ramp from 0.1MPa to 101MPa at a temperature of 37$^\circ$C.  The sample is an off-critical composition mixture of water and 2,6-lutidine emulsified by interfacial silica particles (volume fraction 0.5\%). The frames are recorded every 0.5 seconds; the scale bar is 100 $\mu$m. \label{fig:col1}}
\end{center}
\end{figure}

Figure \ref{fig:col1} shows a series of images of a binary fluid of  water and 2,6-lutidine containing colloidal silica particles, as it undergoes a slow change in pressure from 0.1MPa to 101MPa. The interfaces between the liquid phases are stabilized by the colloidal silica (this is a particle-stabilized emulsion \cite{herzig}). The silica is dyed using the fluorescent dye FITC and a small fraction of the silica particles are dispersed in the continuous phase of the emulsion. The use of fluorescence facilitates imaging deeper into the sample. At the start of the experiment the binary fluid sample is under conditions  (37$^\circ$C, atmospheric pressure) where the two liquids are demixed. On increasing the pressure the liquids begin to remix -- reducing the interfacial tension and hence releasing some of the particles trapped at the interface. The particles are pushed into close contact as the emulsion is destroyed, resulting in some residual clustering in the final frame, even though by this frame the liquids are completely demixed.

\begin{figure}[h!]
\begin{center}
{\rotatebox{0}{{\includegraphics[scale=0.6,clip=true]{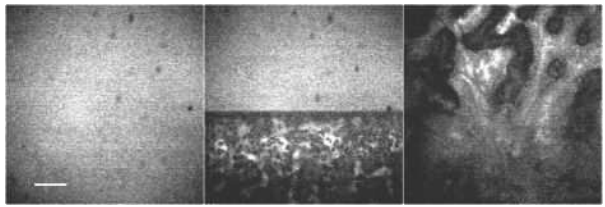}}}}
\caption{Fluorescence confocal microscopy images  of phase separation induced by a pressure quench. The sample is a dispersion of fluorescent silica colloids (volume fraction 0.5\%) in critical composition water-2,6-lutidine. Time increases from left to right in this series of images; the time separation between images is 1.034 seconds; the scale bar is 100 $\mu$m. Each image is 512 x 512 pixels and is produced by rastering a laser at 500 lines per second.  At the start of the experiment the system is mixed: the pressure release causes a demixing transition. \label{fig:col2}}
\end{center}
\end{figure}

We next demonstrate the use of pressure cell 1 for a rapid pressure quench experiment on the water, 2,6-lutidine, colloidal silica system. Here, we start the experiment in the single-fluid region of the phase diagram where the two liquids are mixed (101 MPa and 35$^\circ$C, achieved by first pressurising then warming the sample). At this temperature and at atmospheric pressure the liquids would be in the demixed, two-fluid phase. We can therefore observe the demixing transition by rapidly releasing the pressure; this is achieved by dumping the pressurising brake fluid out of the ram through a solenoid valve.

 The confocal microscopy images captured during the pressure quench experiment are shown as a time sequence in Figure \ref{fig:col2}. Since the images are formed by scanning a laser, time progresses both between frames (left to right) and from the top to the bottom of a frame. The first frame  in Figure \ref{fig:col2} shows a dispersion of particles in the single-fluid phase. The brake fluid is released while the laser is scanning the second frame and this induces a rapid phase separation: a dramatic change in the sample properties occurs over the scanning of five horizontal lines. By the third frame, the phase separation is well developed; the colloidal particles partition into one of the phases which appears brighter than the other. This series of images can be used to estimate the speed of the pressure quench: since the phase transition occurs within the scanning of five lines (at 500 lines per second), we estimate that the quench occurs in 10 ms. We note that the sudden change in pressure under adiabatic conditions results in a concomitant change in the temperature \cite{wong} -- we verified that this change in temperature did not itself cause a phase transition.

\section{Application 2: Bacterial growth under pressure}\label{sec:bacteria}

\begin{figure}[h!]
\begin{center}
\makebox[15pt][l]{(a)}{\rotatebox{0}{{\includegraphics[scale=0.35,clip=true]{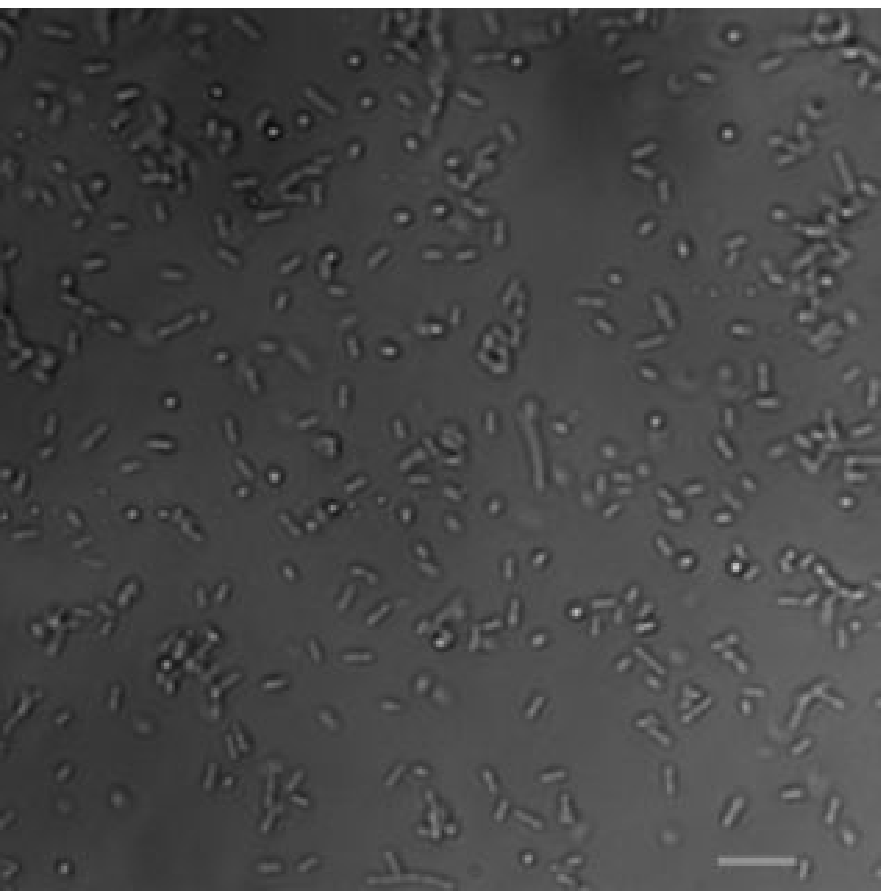}}}}\hspace{0.5cm}\makebox[15pt][l]{(b)}{\rotatebox{0}{{\includegraphics[scale=0.35,clip=true]{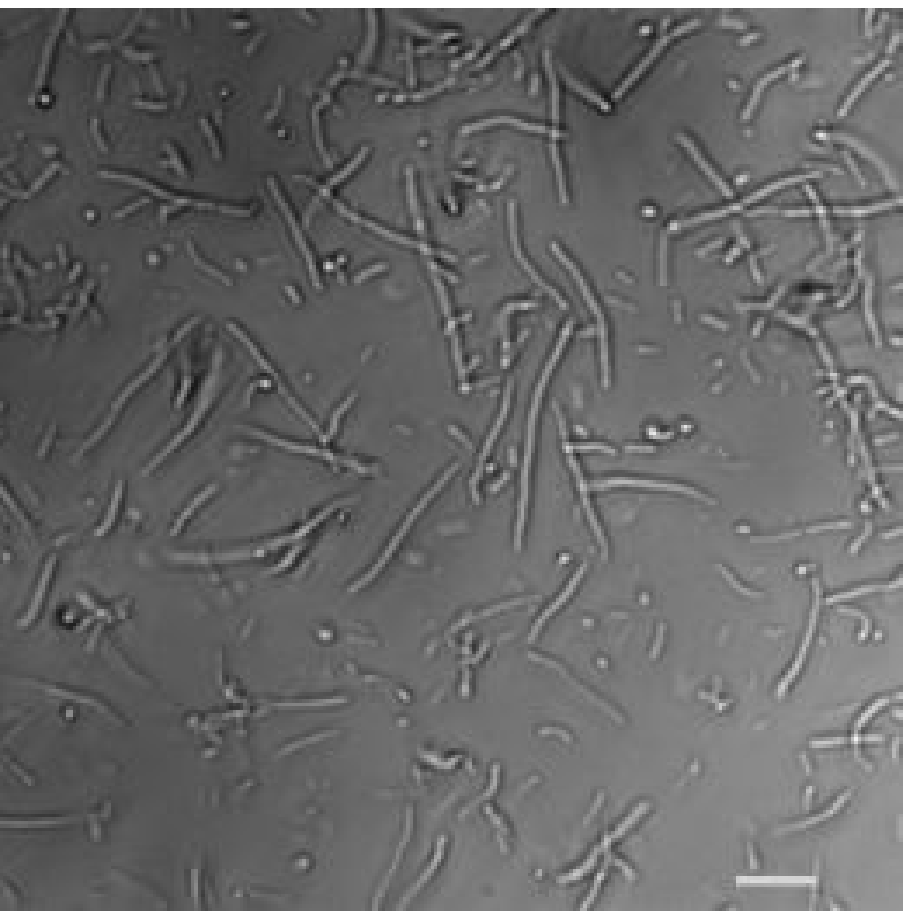}}}}\\\vspace{0.2cm}\makebox[15pt][l]{(c)}{\rotatebox{0}{{\includegraphics[scale=0.35,clip=true]{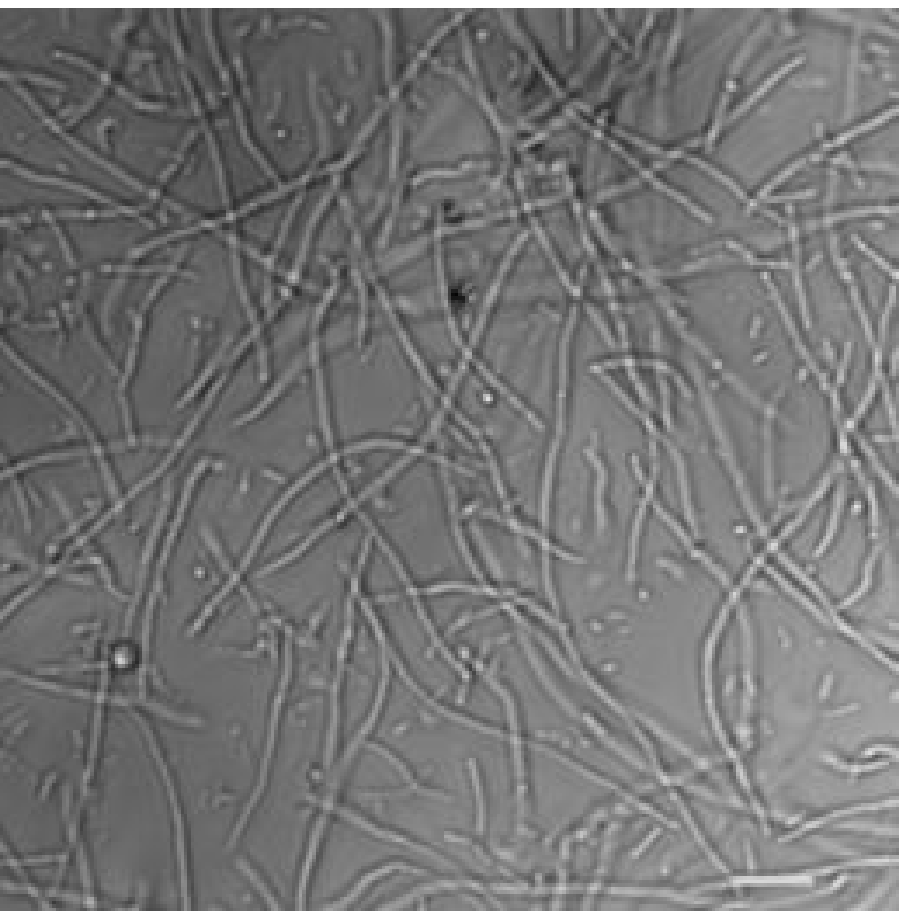}}}}
\caption{{\em{E. coli}} MG1655 growing in pressure cell 2 at 37$^\circ$C at 50MPa (a) 120 minutes, (b) 620 minutes, (c) 20 hours after pressurisation. The scale bar is 10$\mu$m. In these experiments, cells were first grown in a flask in Luria Bertani (LB) medium supplemented with 25mM glucose for 2 hours at 0.1MPa and 37$^\circ$C. The culture was then loaded into pressure cell 2 and left for 30 minutes to allow cells to attach to the window, before being poured off and replaced by fresh LB-glucose medium. The pressure cell was then sealed using the hollow piston and pressurised. Bright-field images were captured every 20 minutes on an inverted Nikon Ti-U microscope with a Sony CoolSnap camera using a 60$\times$ ELWD objective (NA 0.70).
\label{fig:bact1}}
\end{center}
\end{figure}

Hydrostatic pressure is of interest in microbiology in several contexts. Understanding how deep sea microorganisms have evolved to tolerate and even require pressure \cite{bartlett,kato2} has potential applications in biotechnology as well as marine ecology. Applying pressure to  non-pressure-adapted microorganisms  is also of  interest, both as a means of food sterilisation \cite{martin,manas,moussa} and in the general context of microbial stress response \cite{welch,aertsen1,aertsen2,aertsen3,abe}. In all these areas, the ability to image microorganisms in a pressurised microscope cell can provide powerful  insights. For example, microscopy can reveal pressure effects on DNA configuration  \cite{manas,moussa} or protein localisation  \cite{ishii} within the cell. A microscopic approach also allows one to detect heterogeneity among the cellular population (which is believed to be an important factor in  survival of environmental stresses \cite{aertsen_rev,booth}), and to measure single-cell properties such as swimming speeds \cite{eloe}.

\begin{figure}[h!]
\begin{center}
\makebox[15pt][l]{(a)}{\rotatebox{0}{{\includegraphics[scale=0.35,clip=true]{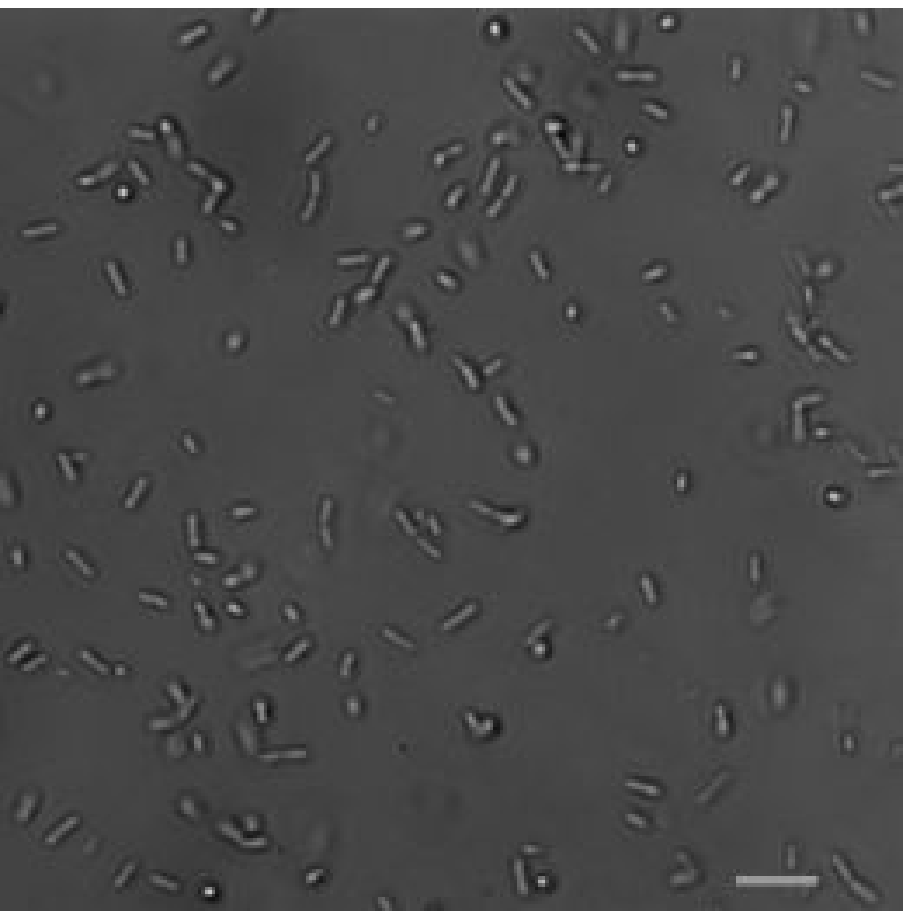}}}}\hspace{0.5cm}\makebox[15pt][l]{(b)}{\rotatebox{0}{{\includegraphics[scale=0.35,clip=true]{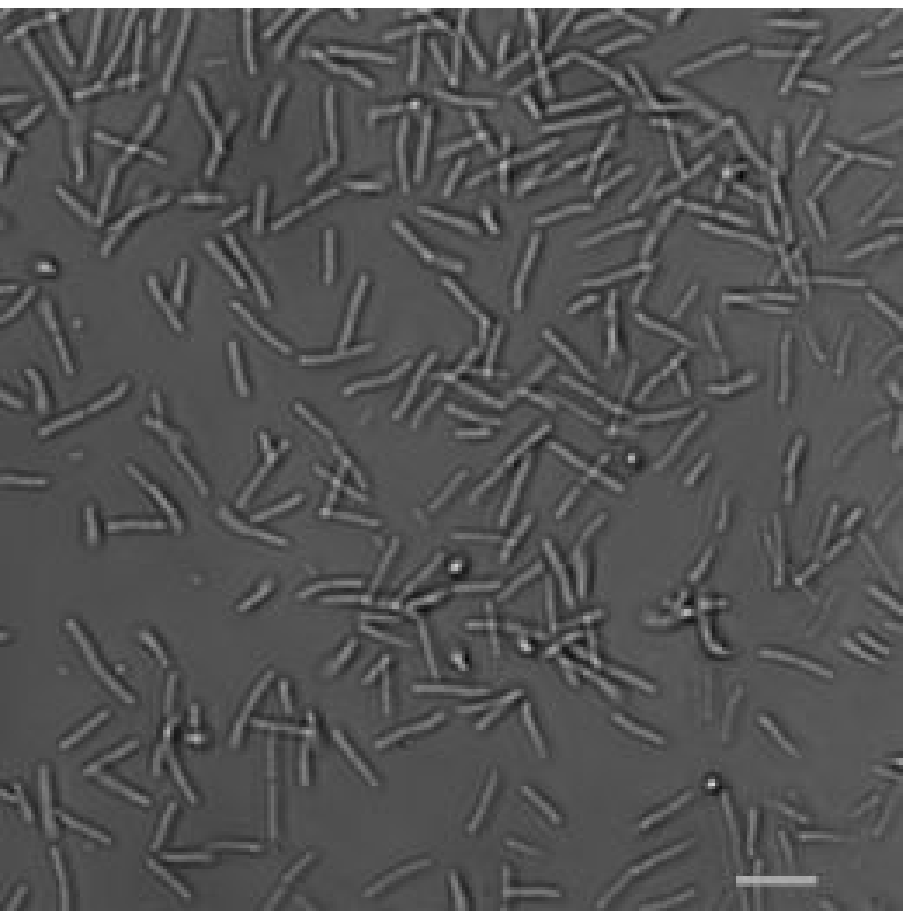}}}}\\\vspace{0.2cm}\makebox[15pt][l]{(c)}{\rotatebox{0}{{\includegraphics[scale=0.35,clip=true]{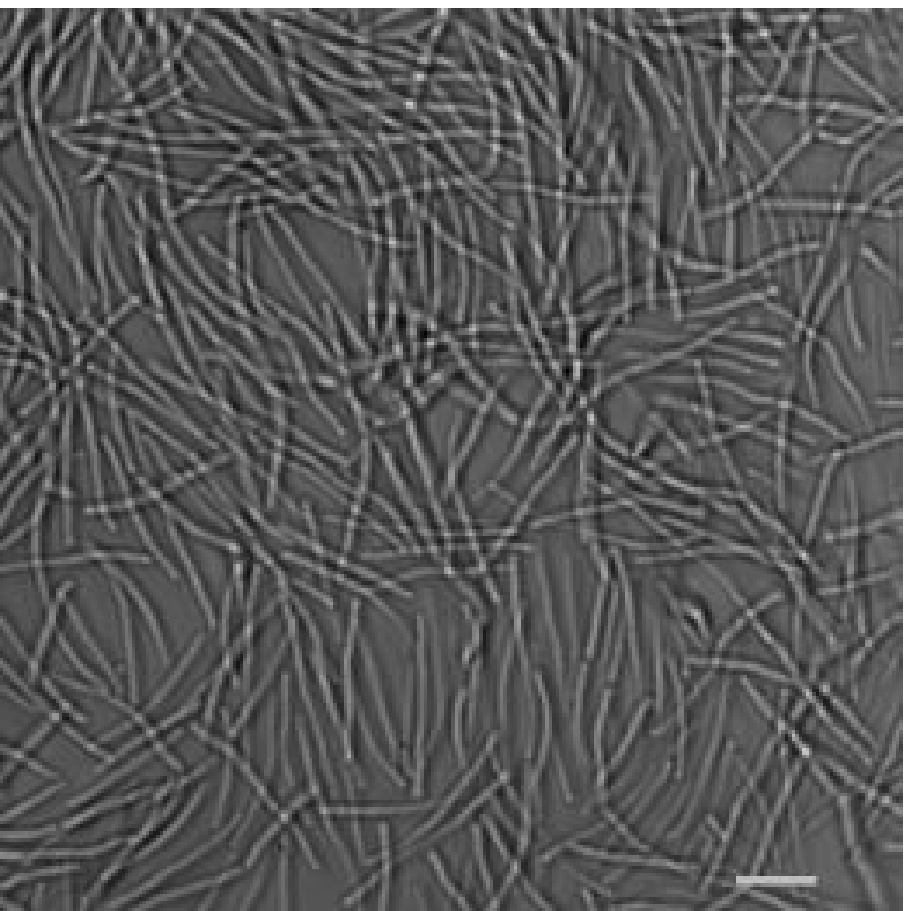}}}}\hspace{0.5cm}\makebox[15pt][l]{(d)}{\rotatebox{0}{{\includegraphics[scale=0.35,clip=true]{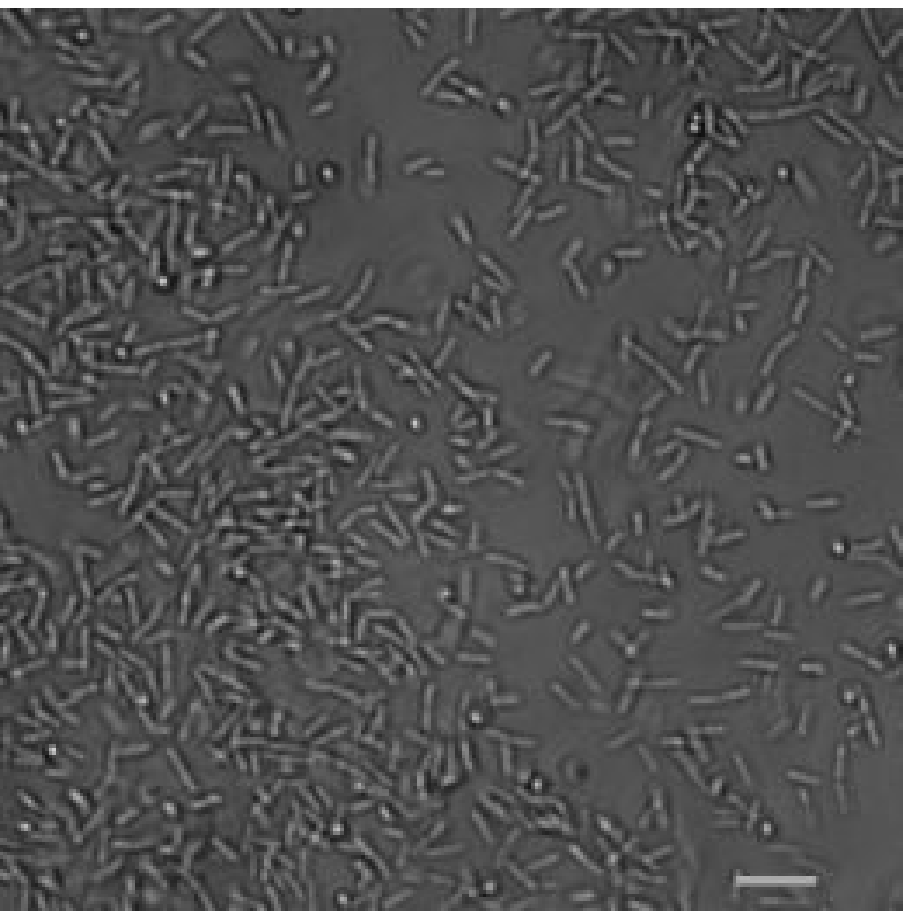}}}}
\caption{{\em{E. coli}} BW25113$\Delta${\em{lon}} growing in pressure cell 2 at 37$^\circ$C at 50MPa (a) 40 minutes, (b) 200 minutes, (c) 400 minutes after pressurisation, and (d) the same strain growing at at 37$^\circ$C and 0.1MPa, imaged 120 minutes after inoculation. The scale bar represents 10$\mu$m. These experiments were performed as described in the caption to Figure \ref{fig:bact1}. Strain BW25113$\Delta${\em{lon}} was obtained from the Keio collection of deletion mutants \cite{keio}. 
\label{fig:bact2}}
\end{center}
\end{figure}

Here, we demonstrate that our system can be used to monitor pressure-associated morphological changes in the bacterium {\em{Escherichia coli}}. {\em{E. coli}} is a non-pressure-adapted organism whose cells are rod-shaped and of typical size 1$\times$2$\mu$m when growing exponentially at atmospheric pressure. The morphological response of {\em{E. coli}} to pressure is   well documented. The details are strain-dependent, but a general picture is that when exposed to pressures up to $\sim$25MPa, {\em{E. coli}} cells continue to grow and divide, but become somewhat elongated compared to those grown at atmospheric pressure \cite{zobell,zobell3,aertsen_book}. At higher pressures, in the range 30-50MPa, cell division is inhibited, leading to filamentous growth \cite{ishii,sato,welch,zobell,zobell3}, while growth is abolished altogether above $\sim$60MPa \cite{zobell,ishii,welch}.  Exposure to pressures above $\sim$150MPa leads to cell death: microscopic studies of {\em{E. coli}} after short-term exposure to such pressures have reported filamentation  \cite{kawarai}, aggregation of cytoplasmic proteins \cite{manas,moussa} and nucleoid condensation \cite{manas,moussa}. 

Here, we present transmitted light images of {\em{E. coli}} cells growing at 50MPa, obtained using pressure cell 2 on an inverted Nikon Ti-U microscope with a 60$\times$ ELWD objective. We compare the behaviour of two {\em{E. coli}} strains: the wild-type strain MG1655 and strain BW25113$\Delta${\em{lon}} \cite{keio} which is unable to produce the Lon protease enzyme.

Figure \ref{fig:bact1} shows {\em{E. coli}} MG1655 cells growing at 50MPa, 120 minutes (a), 620 minutes (b) and 20 hours (c) after pressurisation. The experimental protocol is given in the caption. Filamentation is clearly apparent (whereas the same strain grown in the pressure cell at 0.1MPa shows no filamentation; data not shown). Interestingly, the population is heterogeneous: there is variability in length among those cells that do filament and some cells apparently do not filament at all. We have observed motility of some of these non-filamented cells even after 20 hours growth at 50MPa. A distribution in lengths among filamentous cells grown under pressure has been noted before \cite{zobell3}, but to our knowledge this variability has not been studied in detail.

Figure \ref{fig:bact2} shows equivalent images for the mutant BW25113$\Delta${\em{lon}} \cite{keio}, which is unable to produce the Lon protease. This strain, like MG1655, is able to divide normally when grown at 0.1MPa, but produces filaments when grown at 50MPa. However, in contrast to MG1655, the {\em{lon}} mutant population appears to filament homogeneously, with fewer unfilamented cells being visible. Previous work by Aertsen and Michiels has shown that {\em{lon}} mutants display hyperfilamentation after brief periods of pressurisation at 100MPa, and that this is due to activation of the SOS response pathway \cite{aertsen2,aertsen4,aertsen_book}. Interestingly, however, previous work characterising changes in the {\em{E. coli}} protein \cite{welch} and mRNA \cite{ishii2005} composition during growth at $\sim$50MPa has not shown activation of the SOS pathway. We plan to carry out a more detailed study of the factors governing filamentation at 50MPa in BW25113$\Delta${\em{lon}} versus MG1655 in future work.

\section{Conclusions}\label{sec:concs}
In this paper, we have described a microscope pressure cell system with a modular design, in which parts can be interchanged to allow for a wide range of applications. Our system consists of a pressurized cell and piston, a ram with associated piston, and a temperature-controlled holder. We have designed two versions of the pressurized cell, suitable for pressures in the ranges 0.1-700MPa (cell 1) and 0.1-200MPa (cell 2). We have also developed two versions of the pressure cell piston, one hollow, to allow for transmitted light imaging, and the other solid and more rugged in design, and the ram has been constructed in both hollow and solid versions. The system was calibrated using the ruby fluorescence technique for pressures in the range 200-700MPa, and using a method based on the isotropic-nematic transition of the liquid crystal 5CB in the pressure range 0.1-200MPa. The latter proved a very simple and convenient calibration method.

We have demonstrated our pressure cell system for  two different applications. Our first application was in colloid science. Here, we used pressure cell 1 to image  phase separation in a binary fluid in which the fluid-fluid interfaces are coated with fluorescent colloidal particles. We observed remixing of a phase separated binary fluid as pressure was slowly increased, as well as sudden demixing of a mixed sample on rapid pressure quenching from 101MPa to 0.1MPa. Our second application was in microbiology. Here, we used pressure cell 2 to obtain high resolution images of the growth of {\em{Escherichia coli}} bacteria at 50MPa. We compared the behaviour of two {\em{E. coli}} strains: the wild-type strain MG1655 and strain BW25113$\Delta${\em{lon}}, which is unable to produce the Lon protease enzyme. We observed heterogeneous filamentation of {\em{E. coli}} MG1655 at 50MPa, with marked variability among individual cells in the population. In contrast, the  {\em{lon}} mutant BW25113$\Delta${\em{lon}}, which is hypersensitive to pressure \cite{aertsen4,aertsen_book}, showed more uniform filamentation across the population. 

We plan to use this pressure cell system for further studies in both colloid science and microbiology. In particular, we are interested in the effects of pressure on the interactions between hydrophobic colloids, as models for protein complexation and aggregation, in further characterising the factors governing {\em{E. coli}} filamentation in response to pressure at the single cell level, and in visualising bacterial gene regulatory responses to pressure, using fluorescent reporter proteins. Other applications for this system could include using pressure to trigger phase transitions in complex fluids (as illustrated here for a binary fluid), single cell imaging of microbial killing using short bursts of high pressure and imaging of calcium signalling in mammalian cartilage cells in response to pressure stimuli. Because our system is designed in a modular way, it can easily be augmented with new components as they become necessary for new applications. Components currently under development for use with this system include an additional pressure cell capable of achieving even higher pressures, a very high resolution cell for use at lower pressures, a module to allow for injection of liquids into the cell under pressure and a ``flow through'' module which would allow continuous perfusion of liquid through the pressurised cell.

\acknowledgements
The authors acknowledge valuable discussions with Doug Bartlett, Gail Ferguson, Andrew Hall and Chiaki Kato. This work was supported by EPSRC under grant EP/E030173 and by the SoftComp EU Network of Excellence. SLB and EMH were funded by EPSRC studentships. RJA was funded by the Royal Society and by the Royal Society of Edinburgh. We thank NIG Japan for providing the Keio collection and the Collaborative Optical Spectroscopy, Micromanipulation and Imaging Centre (COSMIC) for use of the confocal microscope.


\end{document}